\documentclass[prb,nofootinbib,twocolumn,superscriptaddress]{revtex4} 

\usepackage{graphicx}
\usepackage{dcolumn}
\usepackage{bm}
\usepackage{threeparttable}
\usepackage{times}
\usepackage{mathptmx}
\usepackage{lscape}
\usepackage{natbib}
\usepackage{amsmath}
\usepackage{amssymb}
\usepackage{braket}
\usepackage{comment}
\usepackage{color}

\def\degree{\kern-.2em\r{}\kern-.3em}

\begin{document}

\title{ Nonequilibrium Bounds for Canonical Nonlinearity Under Single-Shot Work  }

\author{Koretaka Yuge}
\affiliation{
Department of Materials Science and Engineering,  Kyoto University, Sakyo, Kyoto 606-8501, Japan\\
}%

\author{Yutaro Sakamoto}
\affiliation{
Department of Materials Science and Engineering,  Kyoto University, Sakyo, Kyoto 606-8501, Japan\\
}%

\begin{abstract}
{ For classical discrete systems under constant composition (specifically substitutional alloys), canonical average acts as a map from a set of many-body interatomic interactions to a set of configuration in thermodynamic equilibrium, which is generally nonlinear. In terms of the configurational geometry (i.e., information about configurational density of states), the nonlinearity has been measured as special vector on configuration space, which is extended to Kullback-Leibler (KL) divergence on statistical manifold. Although they successfully provide new insight into how the geometry of lattice characterizes the nonlinearity, their application is essentially restricted to thermodynamic equilibrium. Based on the resource theory (especially, thermo-majorization),  we here extend the applicability of the nonlinearity to nonequilibrium states obtained through single-shot work on Gibbs state. We reveal that the extended nonlinearity for nonequilibrium state is bounded from upper and lower by the information about one of the optimal Renyi divergences for equilibrium states in between practical  and linear systems, and  temperature and work. 

}
\end{abstract}


\maketitle

\section{Introduction}
For classical discrete systems with $f$ structural degrees of freedom (SDFs) on given lattice, specifically substitutional alloys under a \textit{constant} composition, the expectation for configuration under a given coordination 
$\left\{ q_{1}, \cdots, q_{f} \right\}$ in thermodynamic equilibrium can be provided by the following canonical average
\begin{eqnarray}
\label{eq:can}
\Braket{ q_{p}}_{Z} = Z^{-1} \sum_{i} q_{p}^{\left( i \right)} \exp \left( -\beta U^{\left( i \right)} \right),
\end{eqnarray}
where $\Braket{\quad}_{Z}$ represents the canonical average, $\beta$ the inverse temperature, $Z=\sum_{i}\exp\left(-\beta U^{\left(i\right)}\right)$ the partition function, with the summation over all possible configurations $i$: e.g., coordinate $q_{k}$ as $k$th multisite correlation function defined by the generalized Ising model,\cite{ce} forming complete basis functions. 
Then the potential energy $U^{\left( k \right)}$ for configuration $k$ is given by
\begin{eqnarray}
\label{eq:u}
U^{\left( k \right)} = \sum_{j=1}^{f} \Braket{U|q_{j}} q_{j}^{\left( k \right)},
\end{eqnarray}
where $\Braket{\quad|\quad}$ denotes the inner product in the configuration space, e.g., $\Braket{a|b}=\rho^{-1}\sum_{k}a^{\left( k \right)}\cdot b^{\left( k \right)}$ ($\rho$ is normalization constant). 
When we introduce two $f$-dimensional vectors of $\vec{Q}_{Z}=\left(\Braket{ q_{1}}_{Z},\cdots, \Braket{ q_{f}}_{Z}\right)$ and $\vec{U}=\left(\Braket{U|q_{1}},\cdots,\Braket{U|q_{f}}\right)$, the former and latter respectively correspond to the configuration in thermodynamic equilibrium and many-body interatomic interactions in the inner-product form.
Subsequently, the canonical average of Eq.~\eqref{eq:can} can be interpreted as a map $\phi_{\textrm{th}}$ of
\begin{eqnarray}
\phi: \vec{U} \mapsto \vec{Q}_{Z},
\end{eqnarray}
which is generally nonlinear. 

In alloy configurational thermodynamics, due to the complex nonlinearity in $\phi_{\textrm{th}}$, many theoretical approaches have been proposed to capture alloy equilibrium properties: e.g.,  The Metropolis algorithm was devised for effective exploration of the configuration space, followed by advanced techniques including the multihistogram method, multicanonical ensemble, and entropic sampling.\cite{mc1,mc2,mc3,mc4} In terms of another aspect to ascertain many-body interatomic interactions, they employ the generalized Ising model, augmented with optimization techniques like cross-validation, genetic algorithms, and regression in machine learning.\cite{cm1,cm2,cm3,cm4,cm5,cm6} 
Although they yield accurate predictions of alloy equilibrium properties, they do not essentially elucidate the nature of the canonical average for alloys as a nonlinear map, which holds particularly true from the perspective of ``configurational geometry'' informed by the density of states in the configuration space (CDOS')  independently determined from thermodynamic variables such as temperature or energy.

To address these issues, we recently introduced a metric for local nonlinearity at a given configuration as a vector field $\vec{H}$ on the configuration space,\cite{asdf,em2} which clarifies that the magnification of $\phi_{\textrm{th}}$ can be quantified by the divergence and Jacobians of the vector field.\cite{bd} We also propose an additional metric for nonlinearity by expanding the concept of $\vec{H}$ to the statistical manifold, which enables the inclusion of further non-local information of nonlinearity, as Kullback-Leibler (KL) divergence.\cite{ig} We observe a strong positive correlation between the averaged partial contribution to nonlocal nonlinearity across all configurations, and the geometric distance in configurational polyhedra (i.e., convex polyhedra determined from correlation functions rage) between practical binary alloys and ideally separable systems in terms of SDFs.\cite{fcc-geom}

Although these works have successfully introduced metrics for the nonlinearity, they do not provide any information about nonequilibrium state, i.e., their original consideration is essentially restricted to thermodynamic equilibrium. To overcome the problem, we first introduce the concept of the nonlinearity for nonequilibrium state as a natural extention from that for thermodynamic equilibrium. Then, based on the resouce theory, we derive bound for the nonequilibrium nonlinearity, which is characterized by the nonlinearity for thermodynamic equilibrium, Renyi divergence between equilibrium and nonequilibrium state, temperature and work. The details are shown below. 

\section{Concept and Derivation}
\subsubsection*{Nonlinearity Measure}
First, we briefly explain the basic concept of local nonlinearity on configuration space as a vector field $\vec{H}$.\cite{asdf, bd}
When we read the configuration as an $f$-dimensional vector of $\vec{q}=\left( q_{1},\cdots, q_{f} \right)$, $\vec{H}$ is given by 
\begin{eqnarray}
\label{eq:asdf}
\vec{H}\left( \vec{q} \right) = \left\{ \phi\left( \beta \right)\circ \left( -\beta\cdot \Gamma \right)^{-1} \right\}\cdot \vec{q} - \vec{q},
\end{eqnarray}
where $\circ$ denotess the composite map and $\Gamma$ is an $f\times f$ real symmetric covariance matrix of CDOS,  $g\left( \vec{q} \right)$ that is fundamentally independent of many-body interatomic interactions and temperature.  
We have shown that $\phi$ exhibits a globally linear map \textit{iff} the CDOS assumes a multidimensional Gaussian form,\cite{ig}
which reveals that the local nonlinearity in $\phi$ can be decomposed into linear and nonlinear contributions, with the former as the invertible map of $\left( -\beta\cdot\Gamma \right)$. Therefore, (i) when $\phi$ is locally linear at configuration $\vec{q}$, $\vec{H}\left( \vec{q} \right)$ takes a zero-vector, and (ii) the image of the composite map $\phi\left( \beta \right)\circ \left( -\beta\cdot \Gamma \right)^{-1}$ is essentially independent of temperature and interatomic interactions. These certainly suggests that $\vec{H}$ can be \textit{a priori} determined based solely on CDOS, $g\left( \vec{q} \right)$. 

Then, we extend the concept of $\vec{H}$ to the statistical manifold to capture additional nonlocal nonlinearity information at  configuration $\vec{q}$. The corresponding nonlocal nonlinearity at a given configuration $\vec{q}'$, $D_{\textrm{NOL}}$, is defined by the KL divergence of $D\left( P^{\textrm{E}} : P^{\textrm{G}} \right)$.\cite{ig} 
The probability distributions $P^{\textrm{E}}$ and $P^{\textrm{G}}$ are given by
\begin{eqnarray}
\label{eq:cdoss}
P^{\textrm{E}} \left( \vec{q} \right)&=& z^{-1}\cdot g\left( \vec{q} \right)\exp\left[ -\beta\left( \vec{q}\cdot \vec{V} \right) \right]   \nonumber \\
P^{\textrm{G}} \left( \vec{q} \right)&=& z^{\textrm{G}}\cdot g^{\textrm{G}}\left( \vec{q} \right)\exp\left[ -\beta\left( \vec{q}\cdot \vec{V} \right) \right]  ,
\end{eqnarray}
where $g\left( \vec{q} \right)$ corresponds to the CDOS of practical system with covariance matrix $\Gamma$, $g^{\textrm{G}}\left( \vec{q} \right)$ corresponds to the CDOS of synthetically linear system, given by multidimensional Gaussian with the same $\Gamma$, and 
\begin{eqnarray}
z &=&  \sum_{\vec{q}}g\left( \vec{q} \right)\exp\left[ -\beta\left( \vec{q}\cdot \vec{V} \right) \right] \nonumber \\
V &=& \left( -\beta\cdot\Gamma \right)^{-1}\cdot \vec{q}.
\end{eqnarray}
Hereafter, we employ the superscript $\textrm{G}$ as a function of the linear system, as defined for $P^{\textrm{G}}$ and $g^{\textrm{G}}$. Note that  $D_{\textrm{NOL}}$ is independent of the temperature and many-body interactions, which is a common characteristic with $\vec{H}$.

\subsubsection*{Setup for Nonequilibrium State}
\begin{figure}[h]
\begin{center}
\includegraphics[width=0.82\linewidth]{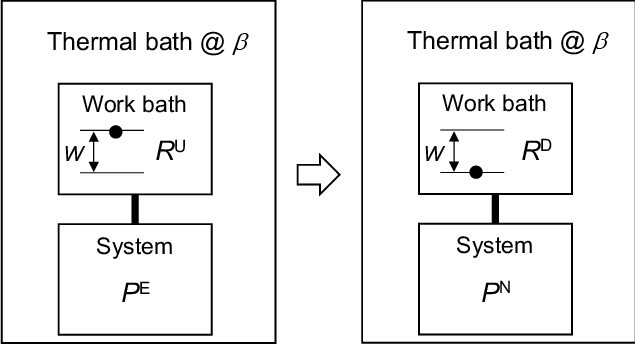}
\caption{ Present setup for preparation of the nonequilibrium state from equilibrium state through the single-shot work. }
\label{fig:setup}
\end{center}
\end{figure}
In order to extend the above concept of the nonlinearity for equilibrium state to nonequilibrium state (NS), we here consider that the NS is prepared from equilibrium state contacting with thermal bath (at inverse temperature $\beta$) and with work bath. The detailed setup is illustrated in Fig.~\ref{fig:setup}, where (i) at the initial time, system takes $P^{\textrm{E}}$ of thermodynamic equilibrium and that for work bath takes the energy level of $+W\ge 0$ (defined as $R^{\textrm{U}}$), and (ii) at the final time, system (with its Hamiltonian unchanged) takes nonequilibrium of $P^{\textrm{N}}$ and the energy level of the work bath of $0$ (defined as $R^{\textrm{D}}$), corresponding to the single-shot work.

Under this setup, we consider the Gibbs preserving map $\Lambda$ for the composite of system and work bath, which provides the  condition for transferability between initial and final states with thermo-majorization: 
\begin{eqnarray}
\label{eq:tm}
P^{\textrm{N}} \otimes R^{\textrm{D}} \prec_{P^{\textrm{E}}\otimes R^{\textrm{E}}} P^{\textrm{E}}\otimes R^{\textrm{U}}, 
\end{eqnarray} 
with 
\begin{eqnarray}
\Lambda \left( P^{\textrm{E}}\otimes R^{\textrm{E}} \right) = P^{\textrm{E}}\otimes R^{\textrm{E}}.
\end{eqnarray}
From the Lorentz curve of the composite system, necessary and sufficient condition for Eq.~\eqref{eq:tm} is given by\cite{tm} 
\begin{eqnarray}
\label{eq:wb}
W \ge \beta^{-1} S_{\infty} \left( P^{\textrm{N}} : P^{E} \right),
\end{eqnarray}
where $S_{\infty}$ denotes Renyi $\alpha$-divergence $S_{\alpha}\left( \alpha \to \infty \right)$ of
\begin{eqnarray}
S_{\alpha}\left( p:q \right) = \frac{1}{\alpha - 1 } \ln \left( \sum_{i} \frac{p_{i}^{\alpha}}{ q_{i}^{\alpha - 1} } \right).
\end{eqnarray}
We note that 
\begin{eqnarray}
\underset{\alpha\to 1}{\lim}S_{\alpha}\left( p:q \right) = D\left( p:q \right). 
\end{eqnarray}

\subsubsection*{Definition of Nonequilibrium Nonlinearity}

To address the nonlinearity for nonequilibrium state, we first divide the canonical average as the following composite of canonical map $\Phi$ and taking avarage for probability distribution $\Braket{\quad}_{1}$: 
\begin{eqnarray}
\phi = \Braket{\quad}_{1} \circ \Phi\left( g \right),
\end{eqnarray}
where
\begin{eqnarray}
\Phi\left( g \right) : \vec{V} \mapsto P^{\textrm{E}}
\end{eqnarray}
under CDOS of $g$.
\begin{figure}
\begin{center}
\includegraphics[width=0.5\linewidth]{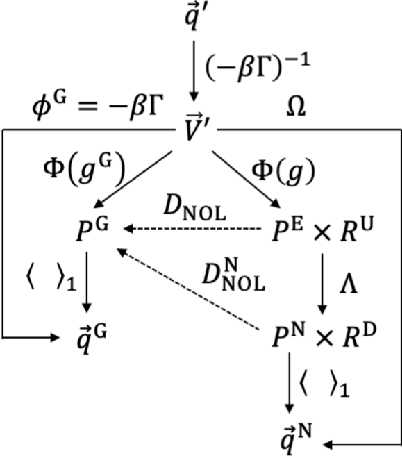}
\caption{ Schematic relationships between nonlinearity for equilibrium state $D_{\textrm{NOL}}$ and nonequilibrium state $P^{\textrm{N}}$ obtained through Gibbs preserving map $\Lambda$. Solid arrows denote taking map, and dashed arrows represents taking difference in probability distributions through (appropriate) divergence.  }
\label{fig:map}
\end{center}
\end{figure}
Based on these maps, we show in Fig.~\ref{fig:map} the relationship between the previously-introduced nonlinearity $D_{\textrm{NOL}}$ for equilibrium state and nonequilibrium state obtained through the Gibbs preserving map $\Lambda$. 
From the figure, in analogy to the equilibrium state, nonlinearity for nonequilibrium state can be introduced as the nonlinear character for map $\Omega$:
\begin{eqnarray}
\Omega : \vec{V} \mapsto \vec{q}^{\textrm{N}},
\end{eqnarray}
where
\begin{eqnarray}
\Omega = \Braket{\quad}_{1} \circ \Lambda \circ \Phi\left( g \right).
\end{eqnarray}
We can now clearly see that when
\begin{eqnarray}
\label{eq:png}
P^{\textrm{N}} = P^{\textrm{G}}
\end{eqnarray}
is satisfied, 
\begin{eqnarray}
\Omega = -\beta \Gamma
\end{eqnarray}
should always hold on. Therefore, Eq.~\eqref{eq:png} is a sufficient condition where $\Omega$ becomes locally linear map in terms of the configurational geometry. 
With these considerations, we here naturally define the nonlinearity for the nonequilibrium state given by
\begin{eqnarray}
D_{\textrm{NOL}}^{\textrm{N}} = D \left( P^{\textrm{N}} : P^{\textrm{G}} \right),
\end{eqnarray}
which has the common end-point for KL divergence of $D_{\textrm{NOL}}$.
The present purpose is therefore to reveal the bound for nonequilibirum nonlinearity based on the information about given $P^{\textrm{E}}$, $P^{\textrm{G}}$ and $\beta W$ (without explicitly using $P^{\textrm{N}}$). 

\subsubsection*{Derivation of Nonequilibrium Bounds}

For the sake of simplicity, hereinafter we introduce followings:
\begin{eqnarray}
P = P^{\textrm{N}}, Q = P^{\textrm{E}}, R =P^{\textrm{G}}, \beta = 1
\end{eqnarray}
and we here assume that $P$, $Q$ and $R$ have the same support.
From Eq.~\eqref{eq:wb} and the monotonicity in Renyi divergence, we first obtain
\begin{eqnarray}
\label{eq:dneq2}
D\left( P : Q \right) \le  W, 
\end{eqnarray}
which can be used to vanish information about $P$. 

One straightforward approach to obtain the upper bound would be applying the max operation in analogy to the Renyi $\infty$-divergence, namely,
\begin{eqnarray}
\label{eq:str01}
D\left( P:R \right) &=& \sum_{x} P_{x} \ln \frac{P_{x}}{R_{x} } \nonumber \\
&=& \sum_{x} P_{x} \ln \frac{P_{x}}{Q_{x} }  \frac{Q_{x}}{R_{x} } \nonumber \\
&=& D\left( P : Q \right) + \sum_{x} P_{x} \ln\frac{Q_{x}}{R_{x} } \nonumber \\
&\le& D\left( P : Q \right)  + \sum_{x} P_{x} \ln\left\{ \max_{k}\left( \frac{Q_{k}}{R_{k} } \right) \right\} \nonumber \\
&=& D\left( P : Q \right) +  S_{\infty} \left( Q : R \right).
\end{eqnarray}
Substituting Eq.~\eqref{eq:dneq2} into Eq.~\eqref{eq:str01}, we obtain the upper bound for nonequilibrium nonlinearity:
\begin{eqnarray}
\label{eq:bd-00}
D\left( P:R \right) \le S_{\infty} \left( Q : R \right) +  W.
\end{eqnarray}

We also show another straightforward approach using Eq.~\eqref{eq:wb}:
\begin{eqnarray}
D\left( P:R \right) &=& \sum_{x} P_{x} \ln \frac{P_{x}}{R_{x} } \nonumber \\
&\le& -S\left( P \right) + \sum_{x} Q_{x} e^{ W}\left( -\ln Q_{x} \right) \nonumber \\
&=& -S\left( P \right) + e^{ W} \left\{ D\left( Q : R \right) + S\left( Q \right)\right\}, \nonumber \\
\quad
\end{eqnarray}
where $S\left( P \right)$ denotes shannon information about $P$. When we employ that $S\left( P \right)$ takes nonnegative by its nature, we obtain the another upper bound:
\begin{eqnarray}
\label{eq:bd-01}
D\left( P:R \right) \le e^{ W} \left\{ D\left( Q: R \right)  + S\left( Q \right)\right\}.
\end{eqnarray}

Then we qualitatively compare the two derived upper bounds of Eqs.~\eqref{eq:bd-00} and \eqref{eq:bd-01}: (i) The common feature is using the bound from the single-shot work as Renyi $\infty$ or KL divergence to vanish the information about $P$. (ii) The former bound is further loosen by approximating the sum of $\ln\left( Q_{i}/R_{i} \right)$ as its max operation, leading to the bound expressed with Renyi $\infty$-divergence of $S_{\infty} \left( Q : R \right)$. (iii) The latter bound is further loosen by completely neglecting the shannon information about $P$, leading to the bound with the exponential factor $e^{ W}$, which is expected to be worse than the former bound especially for high $ W$ and/or $P$ nearly taking random distribution. Eqs~\eqref{eq:bd-00} and \eqref{eq:bd-01} both indicates that higher upper bound can be achieved by increasing $W$ and/or enhancing the difference between equilibrium distribution $Q$ and linear-system distribution $R$. 

Considering these characteristics and the factor for loosening the two upper bounds, further appropriate bound is desired satisfying the following conditions: (i) Avoiding the exponential factor for the work $e^{W}$, (ii) Introducing more sophisticated optimization beyond max operation approximation, from the viewpoint of Renyi $\alpha$-divergence family, and (iii) Providing physical insight to the nonequilibrium nonlinearity.

To this end, we first consider the following:
\begin{eqnarray}
\label{eq:sc}
\sup D(P:R) \  \textrm{s.t.} \  S_{\infty}\left( P:Q \right)\le W, \ \sum_{x} P_{x} = 1,
\end{eqnarray}
which corresponds to constraint optimization problem. 
Here, $S_{\infty}\left( P:Q \right)\le W$ (hereinafter we write $S_{\infty}$-constraint) can be rewriten as
\begin{eqnarray}
^{\forall} x, \ P_{x} - e^{W} Q_{x} \le 0.
\end{eqnarray}
When we consider this problem under given $Q$, $R$ and $W$ based on Karush-Kuhn-Tucker (KKT) condition with the following Lagrangian:
\begin{eqnarray}
\label{eq:L}
L = D\left( P:R \right) - \mu \left( \sum_{x} P_{x} - 1 \right) - \sum_{x} \lambda_{x} \left( P_{x} - e^{W}Q_{x} \right), \nonumber \\
\ 
\end{eqnarray}
necessary condition for the optimal solution (we write as $\hat{P}$) requires the function form of $\hat{P}_{x}$ taking either $e^{W}Q_{x}$ or $cR_{x}$ ($c$ is determined by $\sum_{x}P_{x}=1$): Under the $S_{\infty}$-constraint, upper bound for the nonequilibrium nonlienearity is provided by $D\left( \hat{P}:R \right)$. However, since $\hat{P}$ has the cut-and-paste landscape of $Q$ and $R$ with the constant factor of $e^{W}$ and $c$ on individual microscopic state $x$, the result would not lead to physically insightful bound under the present context of the alloy configurational thermodynamics.

Therefore, we then consider relaxing the $S_{\infty}$-constraint to the following:
\begin{eqnarray}
\label{eq:rc}
S_{r} \left(P:Q\right) \le W \ \left(1<r<\infty\right),
\end{eqnarray}
which we call $r$-constraint. From the monotonicity in Renyi divergence of $S_{\alpha} >= S_{\alpha'}$ for $\alpha > \alpha'$, lower $r$ leads to enhancing relaxation of the original $S_{\infty}$-constraint ($S_{\infty}$-constraint is always the sufficient condition for $r$-constraints). In a similar fashion to Eq.~\eqref{eq:L}, when we employ KKT condition under the $r$-constraint, the optimal $\hat{P}$ condition includes solving $\ln X + a X^{b} + c =0$ type transcendental equation for individual state, which is also out of our present scope. The exception is the $r\to 1$ limit, where the optimal form of the $\hat{P}$ is given by
\begin{eqnarray}
\label{eq:tilt}
\hat{P}_{x} \propto Q_{x}^{\alpha} R_{x}^{1-\alpha},
\end{eqnarray}
which reminds the Renyi $\alpha$-divergence family. Based on these information, hereinafter we focus on the upper bound under the relaxed conditioin of $D\left( P:Q \right) \le W$ (hereinafter we write $D$-constraint). 

We then start from the following Donsker-Varadhan (DV) variational formula for the KL divergence:\cite{dvv}
\begin{eqnarray}
\label{eq:dkl-var}
D\left( p:q \right) = \sup_{\phi \in M_{b}}\left\{ E_{p}\left[ \phi \right] - \ln E_{q}\left[ e^{\phi} \right] \right\},
\end{eqnarray}
where $M_{b}$ is the set of all possible bounded function on support $P$ and $Q$, and $E_{P}$ denotes expectation under $P$. 
The equality condition for Eq.~\eqref{eq:dkl-var} holds when
\begin{eqnarray}
\label{eq:pp}
P_{x} \propto Q_{x} e^{\phi_{x}}
\end{eqnarray}
is satisfied.  It is now clear that when we set the bounded function as
\begin{eqnarray}
\phi_{x} = \left( \alpha -1 \right)\ln \frac{Q_{x}}{R_{x}},
\end{eqnarray}
P defined in Eq.~\eqref{eq:pp} is consistent to that in Eq.~\eqref{eq:tilt}.  Therfore, in order to derive the appropriate upper bound, we set 
\begin{eqnarray}
\label{eq:set}
p &=& P\nonumber \\
q &=& Q\nonumber \\
\phi &=& \left( \alpha - 1 \right) \ln\frac{Q}{R},
\end{eqnarray}
where $\alpha > 1$. 
Then we can transform the variational formula as follows:
\begin{widetext}	
\begin{eqnarray}
\left( \alpha-1 \right)\sum_{i} P_{i} \ln\frac{Q_{i}}{R_{i} }  &\le&   D\left( P:Q \right) + \ln \sum_{i}Q_{i}\left( \frac{Q_{i}}{R_{i}} \right)^{\alpha - 1} , \nonumber \\
\sum_{i} P_{i}\ln\frac{Q_{i}}{R_{i} } = D\left( P:R \right) - D\left( P : Q \right) &\le& \frac{D\left( P:Q \right)}{\alpha - 1 } + \frac{1}{\alpha - 1} \ln\sum_{i}\frac{\left( Q_{i} \right)^{\alpha} }{\left( R_{i} \right)^{\alpha-1} } \nonumber \\
&=& \frac{D\left( P:Q \right)}{\alpha - 1 } + S_{\alpha}\left( Q: R \right), \nonumber \\
D\left( P:R \right) &\le& S_{\alpha}\left( Q:R \right) +  \frac{\alpha}{\alpha-1}D\left( P:Q \right).
\end{eqnarray}
\end{widetext}
Substituting the relaxed single-shot work bound of Eq.\eqref{eq:dneq2} ($D$-constraint) into the last equation, we obtain the upper bound for nonequilibrium nonlinarity:
\begin{eqnarray}
\label{eq:bd-fin}
D\left( P:R\right) &\le& \inf_{\alpha > 1} \left[ S_{\alpha}\left( Q: R \right) + \frac{\alpha}{\alpha-1 } W \right].
\end{eqnarray}
 Now it is clear that (i) the derived bound avoids the exponential factor for work $e^{ W}$, and (ii) the upper bound in Eq.~\eqref{eq:bd-00} is a special case of Eq.~\eqref{eq:bd-fin} with $\alpha\to\infty$, certainly indicating that Eq.~\eqref{eq:bd-fin} strictly provides tighter bound than Eq.~\eqref{eq:bd-00}. 

We note here that the form of upper bound in Eq.~\eqref{eq:bd-fin} can also be obtained in more straightforward way, by considering the maximization of $D\left( P:R \right)$ under $D$-constraint with the Lagrangian relaxation: 
\begin{widetext}
\begin{eqnarray}
\label{eq:lr}
\sup_{\substack{\sum_{x}P_{x}= 1\\ D\left( P:Q \right) \le W} } D\left( P:R \right) &=& \sup_{\sum_{x}P_{x}=1}  \left[ D\left( P:R \right) + \inf_{\lambda\ge 0} \left\{ -\lambda\left( D\left( P:Q \right) -W \right) \right\} \right] \nonumber \\
&=& \sup_{\sum_{x}P_{x}=1}\inf_{\lambda\ge 0} \left[ D\left( P:R \right) -\lambda\left\{ D\left( P:Q \right) - W \right\} \right] \nonumber \\
&\le& \inf_{\lambda\ge 0} \sup_{\sum_{x}P_{x}=1}\left[ D\left( P:R \right) -\lambda\left\{ D\left( P:Q \right) - W \right\} \right]. 
\end{eqnarray}
\end{widetext}
We here introduce the following function:
\begin{eqnarray}
F\left( P \right) = D\left( P:R \right) - \lambda D\left( P:Q \right),
\end{eqnarray}
which should be maximized within Eq.~\eqref{eq:lr} under $\sum_{x}P_{x}=1$: The necesarry condition for the optimal $P^{\star}$ can be provided through the Lagrange multiplier method of
\begin{eqnarray}
L = F\left( P \right) - \mu\left( \sum_{x}P_{x}-1 \right),
\end{eqnarray}
which directly leads to the form of
\begin{eqnarray}
\label{eq:pstar}
P_{x}^{\star} \propto R_{x}^{\alpha'} Q_{x}^{1-\alpha'},
\end{eqnarray}
where
\begin{eqnarray}
\alpha' = \frac{\lambda}{\lambda -1 }.
\end{eqnarray}
We note that $P^{\star}$ in Eq.~\eqref{eq:pstar} has the same function form with $\hat{P}$ in Eq.~\eqref{eq:tilt}. Substituting Eq.~\eqref{eq:pstar} into Eq.~\eqref{eq:lr}, we finally obtain 
\begin{eqnarray}
\sup_{\substack{\sum_{x}P_{x}= 1\\ D\left( P:Q \right) \le W} } D\left( P:R \right) &\le& \inf_{\alpha'} \left[ S_{\alpha'}\left( Q: R \right) + \frac{\alpha'}{\alpha'-1 } W \right],
\end{eqnarray}
which has the same expression with the upper bound in Eq.~\eqref{eq:bd-fin} (proper range of $\alpha'$ should be determined through consistency with the derivation from DV variation). 

With these considerations, we see that Eq.~\eqref{eq:bd-fin} can be one of the most natural and reasonable upper bound for nonequilibrium nonlinearity under single-shot work, without using explicit information about $P$. 
Eq.~\eqref{eq:bd-fin} also provides physical insight into which Renyi divergence for (extended) equilibrium nonliearity $S_{\alpha}\left( Q:R \right)$ provides appropriate bound for the noneqilibrium nonlinearity in terms of the work. Larger work typically requires larger value for optimal $\alpha$, indicating that upper bound for nonequilibrium nonlinearity under large work is dominantly characterized by the higher probability state in equilibrium w.r.t. the linear system, which is futher enhanced by the increase of the work. For the limit of $W\to 0$ where the nonequilibrium state stays near equilibrium state, the former nonlinearity can be well bounded from above by the condition of $\alpha\to +1$, approaching to the KL divergence $D\left( Q:R \right)$ (corresponding to the equilibrium nonlinearity). 

We finally, briefly show that the lower bound for $D\left( P:R \right)$ can be obtained in a similar fashion to obtaining its upper bound, by setting $0\le\alpha <1$ for DV variation based derivation under $D$-constraint, leading to 
\begin{eqnarray}
\label{eq:bd-fin2}
D\left( P:R\right) &\ge& \sup_{0\le\alpha < 1} \left[ S_{\alpha}\left( Q: R \right) + \frac{\alpha}{\alpha-1 } W \right],
\end{eqnarray}
which results in the same function form as the upper bound of Eq.~\eqref{eq:bd-fin}. The derived lower bound can also be obtained by 
directly considering the following:
\begin{eqnarray}
\inf D\left( P:R \right) \ \textrm{s.t.}\ \sum_{x}P_{x}= 1, D\left( P:Q \right) \le W
\end{eqnarray}
with the Lagrange relaxation shown as Eq.~\eqref{eq:lr}. 
Note that for any given work, Eq.~\eqref{eq:bd-fin2} can keep nonnegativity for the nature of KL divergence, $D\left( P:R \right)\ge 0$, by always setting $\alpha = 0$:

\section{Conclusions}
We derive upper and lower bounds for the extended concept of canonical nonlinearity in the nonequilibrium state. The bounds are reasonablly characterized by the Renyi $\alpha$-divergence between equilibrium state for practical and linear systems, and work, where the optimal $\alpha$ depends on the work.

\section{Acknowledgement}
This work was supported by Grant-in-Aids for Scientific Research on Innovative Areas on High Entropy Alloys through the grant number JP18H05453 and  from the MEXT of Japan, and Research Grant from Hitachi Metals$\cdot$Materials Science Foundation.

\end{document}